\documentclass{article}

\usepackage{arxiv}

\usepackage[utf8]{inputenc} 
\usepackage[T1]{fontenc}    
\usepackage{hyperref}       
\usepackage{url}            
\usepackage{booktabs}       
\usepackage{amsfonts}       
\usepackage{nicefrac}       
\usepackage{microtype}      
\usepackage{lipsum}
\usepackage{graphicx}

\title{Information Theory as a Means of Determining the Main Factors Affecting the Processors Architecture}

\author{
  Anton Rakitskiy\\
Novosibirsk State University\\
Institute of Computational Technologies of the Siberian Branch of the RAS\\
Siberian State University of Telecommunication and Information Sciences\\
  \texttt{rakitsky.anton@gmail.com} \\
   \And
 Boris Ryabko\\
Institute of Computational Technologies of the Siberian Branch of the RAS\\
Novosibirsk State University\\
  \texttt{boris@ryabko.net} \\
}

\begin{document}
\maketitle

\begin{abstract}
In this article we are investigating the computers development process in the past decades in order to identify the factors that influence it the most. We describe such factors and use them to predict the direction of further development. To solve these problems, we use the concept of the Computer Capacity, which allows us to estimate the performance of computers theoretically, relying only on the description of its architecture.
\end{abstract}

\keywords{computer capacity \and processors architecture \and information theory \and performance \and processors characteristics}

\section{Introduction}

Currently, there is a wide variety of processors and computing devices with them in information technologies. The main purpose of this work is to analyze the process of their development and to identify the factors that can be used to predict and control the "evolution" of processors. To achieve this, we propose to use the Computer Capacity characteristic. The main concept of the Computer Capacity characteristic was originally proposed in 2012 \cite{perf_eval}. There was shown that using this characteristic we can estimate the performance of any computer theoretically, based solely on the description of its architecture. In other words, the evaluation of the Computer Capacity does not require a working model of the investigated computer. 
In the following papers \cite{jcsc1, jcsc2} it was shown that the presented characteristic is consistent with the results of generally accepted benchmarks \cite{passmark,coremark,top500,overall} for processors of different types. In our works we investigated the majority of processors types, such as Intel, AMD, ARM, MIPS and graphical processors as well as supercomputers built on them. The results of presented researches have shown that the Computer Capacity characteristic is fairly accurate estimation for the computers performance and, due to the possibility of theoretical evaluation, it can be used at the computer development design stage. 

In addition, in our last work\cite{evolution}, the characteristic was successfully applied to analyze the evolution of Intel processors. In the course of this study, an evolutionary series of Intel processors over the past 20 years was considered and parameters which have a non-linear effect on processor performance were identified. It was shown that in practice, these parameters were increased by manufacturers when moving from the old processor to the new. It should be noted that the Computer Capacity allows not only to reveal such parameters, but also to quantify their influence.

In this paper, we show how the Computer Capacity allows us to identify the main factors that determine the trends in the development of processors architectures. The main parameters that determine the development of architectures in the past and in the future are highlighted. Moreover, the Computer Capacity allows not only to highlight these factors, but also to quantify their impact on various architectures. All such factors can be divided into:
\begin{itemize}
\item The factors whose influence is obvious, such as number of cores, clock frequency, etc.
\item The factors whose influence is not so obvious are the number of threads (by threads we mean the processor's ability to execute independent instructions in parallel), the sizes and the access times of different type of memory, and the set of instructions.
\end{itemize}
In this article, we focus on the factors of the second group and show how the future development will depend on them. 

\section{The main concept of the Computer Capacity characteristic}

In this section we briefly describe the main definitions and a summary of the theory to understand the concept of the Computer Capacity. The entire theoretical basis with the description and proof of the theorems can be found in \cite{perf_eval}.

In general, a computer can be represented as the set of instructions $I$ and the memory $M$ to which these instructions address. An instruction $z\in I$ is a combination of the instruction name and the values of its operands. It means that, for example, instructions "mov ax, bx" and "mov dx, ax" are different and both are included in set $I$. We consider computer task $Z$ as the sequence of instructions $Z=z_1,z_2,..., z_n \in I$. Let us note that if there is a loop in task which is repeated $k$ times the instructions from the body of this loop are included $k$ times in $Z$. We denote the execution time of instruction $z \in I$ as $\tau(z)$ and the execution time of computer task $Z=z_1,z_2,...,Z_n$ as the sum of instructions execution times $\tau(Z)=\sum_{i=1}^{n}{\tau(z_i)}$.

Let us consider the number of all possible computer tasks which execution time equals to $T$ as $N(T)=|{Z:\tau(Z)=T}|$. In \cite{perf_eval} there was showed that this number grows exponentially as the function of time. If there are $N_1$ tasks with execution time 1 minute, there are $N_1^2$ tasks with execution time 2 minutes, and correspondingly there are $N_k\approx N_1^k$ tasks which are executed in $k$ minutes. In this way $N(T)\approx 2^{CT}$, where $C$ is the measure which we call the Computer Capacity. So the Computer Capacity can be considered as follows:
\begin{equation}
\label{eq1}
C(I)=\lim_{T \to \infty}{\frac{\log{N(T)}}{T}}.
\end{equation}

The main task here is how to estimate the value of $C(I)$ from (\ref{eq1}). The direct calculation of this limit is impossible, but there exist the method of calculation $C(I)$ in combinatorial analysis. Here we consider the set of instructions $I$ as an alphabet and assume that all computer tasks $Z$ are words over this alphabet. We will consider that all computer tasks can be executed, i.e. all words over the alphabet $I$ are possible. In this way it allows us to estimate the upper bound of the Computer Capacity, because for any computer the set of its permissible tasks is the subset of all possible tasks.  In addition we assume that all execution times are integers and their greatest common divisor is 1. We should clarify that this assumption is valid for the majority of processors: instructions execution times are mainly measured in processor cycles so they are integers; there is at least one instruction $z$ with $\tau(z)=1$, so the greatest common divisor is also equal 1. The way of estimation of the capacity was suggested by C. Shannon \cite{shannon}, who showed that the capacity $C(I)$ is equal to the logarithm of the largest real solution $X_0$ of the following characteristic equation:
\begin{equation}
\label{eqCC}
X^{-\tau(z_1)}+X^{-\tau(z_2)}+...+X^{-\tau(z_n)}=1,
\end{equation}
where $n$ is the size of $I$. In \cite{perf_eval} it was also proved that the Computer Capacity of multi-core processing unit is defined as the sum of Computer Capacities of the cores. As shown in \cite{perf_eval}, if a computer contains $n$ computational cores and their Computer Capacities are $C_1,C_2,...,C_n$, respectively, so the Computer Capacity of such computer will be: \begin{equation}\label{eqSum}
C_{comp} = \sum_{i = 1}^{n}{C_i}. 
\end{equation}
If all cores are the same, i.e. $C_{core} = C_1 = C_2 = ... = C_n$, the formula can be written as $C_{comp} = n\times C_{core} $.

\subsection{The example of evaluating the Computer Capacity}
To show in practice how to calculate the Computer Capacity let us describe a simple computer with 8 registers, 16 memory cells of level-1 cache and 256 memory cells of RAM. The list of instructions for this computer consists of: $mov r r$, $mov r m$, $add r r$, $mul r r$. Here, operand $r$ means addressing the register and operand $m$ means addressing the memory cell. The memory in the presented computer is arranged as follows: first, the instruction accesses the cache memory and if the required memory cell is not found there, then it accesses the RAM. Let us consider cache-memory access time as 1 clock cycle and the RAM access time as 5 clock cycles. The execution times of instructions are 1,1,2 and 5 cycles in accordance with the list presented above. As mentioned above, the instruction set includes not only the name of the instruction, but also the value of its operands. Also we suppose that all combinations of operands values are possible, so there are $8\times 8=64$ variants of instruction $mov r r$. All of these variants are included in the equation (\ref{eqCC}) as term ${64}/{X^1}$. Similarly the instruction $add r r$ is included in the equation as ${64}/{X^2}$ and the instruction $mul r r$ as ${64}/{X^5}$. Consider separately the instruction $mov r m$. When adding this instruction there are two cases: if instruction addresses cache memory and if the memory cell is not found in cache and instruction addresses RAM memory. In the first case there are $8\times 16 = 128$ different variants with the execution time equals to the sum of execution time of instruction and the cache-memory access time $1+1=2$, so we include this in equation (\ref{eqCC}) as term ${128}/{X^2}$. In the second case the number of different variants of instruction is $8\times 256=2048$ and the execution time is the sum of instruction execution time, cache-memory access time and RAM access time $1+1+5=7$. So it included in the equation as ${2048}/{X^7}$. Now we can construct the equation (\ref{eqCC}) as:
\[
\frac{64}{X^1} + \frac{64}{X^2} + \frac{64}{X^5} + \frac{128}{X^2} + \frac{2048}{X^7}=1
\]
After using the bisection method we get the value of largest real solution $X_0=66.871$ and $C(I)=\log_2{X_0}=6.06$ bits/clock. 

\section{The analysis of the characteristic equation}

In previous works, we showed that the results of applying the method described above correlate well with the results of generally accepted benchmarks. In the appendix A, you can see graphs with the results of the comparison and links to detailed descriptions of equations for real processors of different architectures. In this section, we show how changes in certain parameters of the processor architecture affect its performance, and how this relates to the method of evaluating the Computer Capacity and, in particular, with the peculiarities of constructing the characteristic equation.

In \cite{evolution}, an evolutionary series of Intel processors was investigated, and as a result, it was shown that during the development of Intel processors, manufacturers changed precisely those of the processors architecture complex parameters that affected the Computer Capacity the  most. Such parameters are the number of internal registers of the processor, the number of fast instructions and their types. In addition, it was possible to estimate the impact of changes in these parameters quantitatively, and this assessment coincided with practical data. For example, we showed that if the Wolfdale processor increases the number of internal registers 10 times and adds 16 instructions with three register operands whose execution time is 1 clock cycle, the growth of the Computer Capacity will be \~47.5\%. And the next processor in Intel’s evolutionary series, Ivy Bridge, was changed in exactly this way, the number of registers increased to 160 from 16 and 10 new instructions of the specific type were added (besides, major changes were made to the list of instructions as a whole). The growth of the Computer Capacity was \~53\%. It is interesting to note that changing other parameters, such as cache and RAM and access time to this memory (of course, changing the access time was considered in real terms, without bringing it to instant access, like registers), had almost no influence on the value of the Computer Capacity. What is important is the fact that, in practice, these values also remained unchanged.

\subsection{Dead-end branch}
Here we show in the example how the Computer Capacity could be useful to developers. In this example, it can be argued that the processors development aims to increase the Computer Capacity and if such an increase does not occur then a dead-end may occur. The processor Pentium III was presented in the beginning of 1999 and its successor Pentium IV was presented in November of 2000. It is important to note that both processors based on architectures with different principles. The main task for a new one processor was to increase the maximal clock rate.
 
We have estimated the values of the Computer Capacity for Pentium III and Pentium IV processors and they were equal to $42.021$ and $39.657$ bits per clock cycle respectively. This suggests that with equal clock rates the Pentium III processor can perform more tasks than the Pentium IV processor. Based on this we can assume that the development went the wrong way. Nevertheless, Pentium IV processors developed and were released for several more years, until they reached the peak of their performance. After that Intel Core architecture appeared, which was based on the Pentium III, and currently, modern Intel processors to a certain extent are the heirs of this architecture. Based on the above, we assume that the Pentium 4 processors can be considered as a dead-end branch of evolution. And we shown how the appearance of such a branch could have been avoided by using in the development the Computer Capacity characteristic which clearly showed the predominance of the first processor over the second.

\section{Factors affecting processors architectures}

In this section, we highlight and consider the key factors affecting the development of processors. Here we do not consider the factors whose impact on performance is obvious, and focus on the description of factors with not so obvious effect. First we need to list these factors:
\begin{enumerate}
\item The factors that linearly affect the performance of processor. There is only one such factor, the number of so-called threads. By thread we mean the property of a pipeline to simultaneously execute several unrelated instructions. The number of threads is the maximum number of instructions that the processor pipeline is capable to execute at a time. And it is important to note that such threads use the same memory, i.e. increasing the number of threads does not require an increase in memory. 
\item The factors with non-linear impact on the performance. These factors include the amounts of different types of memory, the access times of these memory types and the set of instructions. These factors have a tendency to rapid saturation, so it is necessary to investigate, among other things, their growth potential.
\item The factor of the little effect of slow instructions on the performance. The concept of the Computer Capacity shows us that the longer the instruction execution time, the less its influence on the Computer Capacity, and this influence decreases exponentially. It is important to note that this factor is directly related to the previous one and moreover, follows from it.
\end{enumerate}

\subsection{Effect of using shared memory}

Formula (\ref{eqSum}) allow us to understand how the number of cores affects the Computer Capacity. But, in addition, the same formula is applicable to so-called threads. Let us consider this characteristic in more detail. The processor pipeline consists of stages, each of which performs its specific task (instructions decoder, renaming registers, execution blocks, etc.). All stages are characterized by the number of instructions which they can execute in parallel (let us call this the throughput of the stage). By the number of processor threads we will mean the minimum value of throughput among all stages of the pipeline (obviously, it is impossible to execute more instructions in parallel than on the stage with minimum throughput).

Threads are independent sequences of instructions (for example, if there is the instruction which loads data from a memory cell into a register, and the second instruction which adds this register to another one, they cannot be executed in parallel, and therefore they fall into dependency chain) which can be executed in parallel, distribution of instructions on chains of dependencies is made by the processor automatically. In fact, in the context of the Computer Capacity, we can assume that these threads are independent processor cores with shared memory (such as registers, cache-memory and RAM). As mentioned earlier, we estimate the upper bound of the Computer Capacity, and the set of all pairs of instructions sequences which can be executed in parallel, obviously, is a subset of the set of all possible pairs of instructions sequences. Increasing the number of threads is an extremely profitable solution from the manufacturer’s point of view, since all threads inside the kernel use the same registers and memory, but adding a thread affects the performance of the processor, ideally, just like adding a new computational core. 

In part, the influence of this factor can be observed in graphics processors \cite{GPU}. Despite the fact that the number of computing cores is increased in graphics processors, these cores are inherently close to the threads we described above. For example, the typical NVIDIA GPU comprises of a set of Streaming Multiprocessors (SM) which share the level-2 cache and DRAM. Each SM, in turn, comprises of several Stream Processor (SP) cores which share the level-1 cache memory and the register file. Thus, we can match stream processors and threads because of the fact that both of them, according to our theory, are reduced to computational cores with shared fast memory. And it is interesting especially due to the fact that modern GPUs consist of thousands of SPs.
\subsection{Logarithmic saturation effect}

The concept of the Computer Capacity allows us to understand why changing different parameters affects the processors performance differently, and, moreover, quantify this effect. First, note that the value of the Computer Capacity is the logarithm of the largest positive root $\log{X_0}$ of the equation (\ref{eqCC}). From mathematical analysis it is well known that when the $x$ parameter is changed, the function will change as follows $\ln{(x+\Delta x)}-\ln{x}\sim {\Delta x}/{x}$. 
For example, the value of (\ref{eqCC}) solution for Intel Wolfdale processor equals $X=13007226$ and its Computer Capacity value is $C=\log_2{X}=\log_2{13007226}\approx 23.632$. So when we increase ten times the number of vector registers in this processor the value of solution becomes $X=1472312547$ and the Computer Capacity is $C=\log_2{1472312547}\approx 30.455$ which fully corresponds to the theory described above.   
It is important to note that the effect of an increase in the number of instructions or memory cells (including the number of registers) also has a non-linear effect on the result of solving the equation (\ref{eqCC}). All the above suggests that the considered here non-obvious parameters have a clear tendency to saturation.

Since accessing registers does not imply a delay, an increase in their number will affect almost all the terms, in the denominator of whose exponent is equal to 1 and 2, those that most strongly influence the value of the Computer Capacity. And if in the instruction two operands are registers, then with an increase in the number of registers 2 times, the numerator will increase 4 times, and if there are three operands, then, respectively, 8 times. In \cite{evolution}, it was shown that with a tenfold increase in the number of integer and vector registers, the value of the Computer Capacity of Intel processors increases by more than 30\%. The same increase in the number of internal registers can be observed in the transition from Wolfdale processors to Ivy Bridge processors. Unfortunately, to get a similar growth, developers need to increase the current number of registers 10 times, but in the Skylake architecture there are already 180 integer and 168 vector internal registers. It will be problematic to achieve this with the existing element base, so we can talk about the approaching saturation.

One of the most difficult factors to evaluate is the number of instructions. There are several ways, the development of new instructions with a large number of operands (as, for example, it was done in Intel processors, when fast instructions with three register operands were added), or the addition of new instructions of the existing type. For example, ARM followed the path of a significant increase in the number of instructions, which can be seen by comparing the architecture of the Cortex-M3 and Cortex-A57 processors. In \cite{arm_doc} you can find a detailed description of all architectures of ARM processors, and the link \cite{link} contains transformed instruction sets, on the basis of which we built the characteristic equation (\ref{eqCC}) for these processors. In these converted lists (they are presented in a special format, where each instruction is divided into terms depending on its execution type), the description of the processor M3 contains 218 instructions, while the list for A57 consists of 1877 instructions. If you look at the equations, you can see that the numerator in the first term (where the degree is 1) in A57 equation (\ref{eqA57}) is three times greater than the numerator of the first term in M3 equation (\ref{eqM3}). At the same time, the Computer Capacity of Cortex-M3 $ C(I_{M3}) = 25.2 $ bits / clock increased to $ C(I_{A57}) = 29.79 $ bits / clock for Cortex-A57, i.e. only 18\%. Nevertheless, the A57 processor significantly exceeds the performance of the M3, primarily due to the fact that the number of threads was increased from 1 to 3, which allowed it to get closer to modern Intel processors (comparing with Intel processors using benchmarks and computing capacity can be found in the appendix). In addition, the development of new instructions is a very time-consuming process, so in this direction we can also note the approaching saturation.

\begin{eqnarray}
\label{eqM3}
\frac{38553518}{X^{1}} + \frac{143623384}{X^{2}} + \frac{35360612}{X^{3}} + \frac{35568868}{X^{4}} + \frac{2142526}{X^{5}} + \frac{1321936}{X^{6}} +\nonumber \\ 
\frac{2239104}{X^{7}} + \frac{2922720}{X^{8}} + \frac{3094292}{X^{9}} + \frac{2579488}{X^{10}} + \frac{1687168}{X^{11}} + \frac{857248}{X^{12}} + \frac{327600}{X^{13}} +\nonumber \\ 
\frac{92400}{X^{14}} + \frac{18000}{X^{15}} + \frac{2160}{X^{16}} + \frac{120}{X^{17}} = 1
\end{eqnarray}

\begin{eqnarray}
\label{eqA57}
\frac{928619266}{X^{1}} + \frac{358487379}{X^{2}} + \frac{72121984}{X^{3}} + \frac{279500883}{X^{4}} + \frac{167809692}{X^{5}} + \frac{31454404}{X^{6}} +\nonumber \\ 
\frac{3735614}{X^{7}}+ \frac{25589699}{X^{8}} + \frac{21567538}{X^{9}} + \frac{7511822}{X^{10}} + \frac{102413}{X^{11}} + \frac{131072}{X^{12}} + \nonumber \\ 
\frac{65536}{X^{14}} + \frac{65536}{X^{15}}=1
\end{eqnarray}

\subsection{Low effect of slow instructions}
As was shown, the amount of memory affects the value of the numerators in the addends, which are formed by instructions working with memory. Let there be an instruction $instr m$ whose execution time is 1 clock cycle, while the computer has a 1-level cache memory with size $L_1$, a cache memory of the 2nd level with size $L_2$ and RAM with size $M$ memory cells. The memory access time is denoted by $\tau(L_1), \tau(L_2), \tau(M)$, respectively. Then such an instruction forms three components:
\[\frac{L_1}{X^{1+\tau(L_1)}},\frac{L_2}{X^{1+\tau(L_1)+\tau(L_2)}},\frac{M}{X^{1+\tau(L_1)+\tau(L_2)+\tau(M)}}.\]
If we increase the cache size of the first level 2 times, the numerator in the first term becomes ${2L_1}/{X^{1+\tau(L_1)}}$. However, it is important to note here that, for example, in Intel processors, the access time to the 1st level cache is 3-4 clocks, and therefore the denominator will have a 4-5 exponent value. It was shown in \cite{evolution} that the processor has a sufficient number of instructions forming terms with exponent 1 or 2, and they have a much greater influence on the solution of the equation (\ref{eqCC}), and, therefore, even a tenfold increase in 1-level cache memory practically does not affect the final value of the Computer Capacity. In \cite{evolution}, we showed that, for Intel processors, a change in the cache sizes of all levels, as well as a change in the size of RAM, does not affect the Computer Capacity, and therefore does not make much sense to the performance. And, as practice has shown, manufacturers really did not change the volume of these types of memory, that is, saturation was achieved. Such a trend could have been significantly reduced by the access time to this memory, however, unfortunately, the technologies used in the investigated processors do not allow this to be done yet. We need to note that such effect of slow instructions corresponds to logarithmic factor, and, in fact, well illustrates the essence of the saturation effect.

\section{Conclusion}

We have shown how the concept of the Computer Capacity helps us to identify and, moreover, to quantify the factors which influence the development of processors. In this paper, we have identified the factors in computer development whose influence on the performance is not so obvious. We divided these factors into two groups, with linear impact on the performance and with the impact close to logarithmic. Most of the parameters of the second group have already reached saturation, but among them are those who have not reached the growth limit. These parameters differ for different processors, for example, ARM processors have potential to grow the number of registers, while Intel processors are already close to the limit of this value (it will be quite problematic to increase their number tenfold). In its turn, the factors from the first group most evident in GPUs, but in fact they are used in all modern processors. It is important to note that the change in the element base of the processors also agrees well with the presented theory. For example, increasing the speed of access to memory will affect the denominators in the characteristic equation and will undoubtedly increase the Computer Capacity, and we have shown how to estimate this increase. 

In previous years, manufacturers implicitly took these factors into account, but unconsciously, based on experiments, benchmarks and their own intuition. It seems to us that explicit consideration of the Computer Capacity and corresponding equation (\ref{eqCC}) as well as described factors can significantly effect the development of processors. Here we can give some conditional analogy with the "evolution" of optical devices. Obviously, the "evolution" did not know what the focal length, refractive index and other characteristics of the optical systems. In the animal world, vision organs developed unconsciously obeying the main optic characteristics (such as the focal length of the lens, which is essentially the basis of the eye) and the laws of optics, but after the advent of optical devices, their development took place with an explicit account of these laws, which, it seems to us, has greatly accelerated this process. We hope that the laws described above, as well as the concept of the Computer Capacity in general, can accelerate the development of processors, as well as devices in which they are used.

\vspace{6pt} 

\section*{Appendix}
Here we present some results of comparing the Computer Capacity with benchmarks for processors with different architectures. Considering the difference in units of measurement, it is not possible to directly compare our characteristic with benchmarks, so a relative comparison has been made. Figure A1 shows how the performance of different Intel processors varies relatively to the ARM Cortex-A57 processor. The comparison presented for single core of each processor, the benchmarks results are taken from \cite{armbench}. All the characteristic equations and calculation programs can be found in \cite{link}.

\begin{figure}[ht]
	\includegraphics{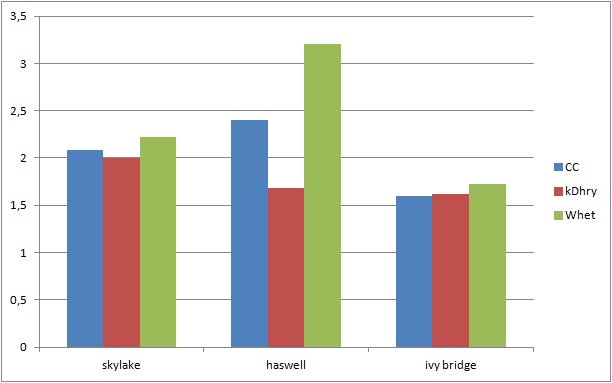}
	\caption{The results of processors comparison}
	\label{fig1}
\end{figure}
\vspace{6pt}
\bibliographystyle{ieeetr}
\bibliography{computer_capacity}





\end{document}